\documentclass[conference]{IEEEtran}
\ifCLASSINFOpdf
\else
\fi
\usepackage{array}

\usepackage{verbatim}
\usepackage{amssymb,amsmath}
\usepackage{url}
\usepackage{mycrypto-ieee}
\usepackage{ifthen}
\usepackage{listings}
\numberwithin{equation}{section}

\usepackage{url}

\begin{document}
%
\title {Multi-user protocols with access control for computational privacy in public clouds}

\author{\IEEEauthorblockN{Sashank Dara\IEEEauthorrefmark{1}
\IEEEauthorrefmark{2}}
\IEEEauthorblockA{\IEEEauthorrefmark{1} Cisco Systems India Pvt Ltd,}
\IEEEauthorblockA{\IEEEauthorrefmark{2} International Institute of Information Technology, \\ Bangalore, India}}


%


\maketitle

\begin{abstract}
\emph{Computational privacy} is a property of cryptographic system that ensures the privacy of data being processed at an untrusted server. Fully Homomorphic Encryption Schemes (FHE) promise to provide such property. Contemporary FHE schemes are suited for applications that have single user and server. In reality many of the cloud applications involve multiple users with various degrees of trust and the server need not necessarily be aware of it too.
We present a \emph{Complementary Key Pairs} technique and protocols based on that to scale any generic FHE schemes to multi user scenarios. We also use such technique along with FHE to show how \emph{attribute based access control} can be achieved while server being oblivious of the same. We analyze the protocols and their security. Our protocols don't make any assumptions on how FHE scheme itself works.
\end{abstract}


%
\IEEEpeerreviewmaketitle

\section{Introduction}
Cloud computing is well defined and standardized \cite {mell2011nist}. The major hindrance for cloud adoption is still  \emph{computational privacy}. It is a property of cryptographic system that ensures the privacy of data being processed at a remote untrusted server.
 
Informally in a cloud computing setup, \emph{Alice} has inputs \emph{\{$x_i$\}} to a function \emph{f} to which she needs results, but lacks enough resources to compute the output. \emph{Sally}, a service provider, has resources to execute \emph{f(x)} and is willing to let \emph{Alice} use them may be for a price. Similar to any Cloud Service Provider like Google's Compute Engine, Amazon's WebServices  etc. \emph{Alice} is concerned though that \emph{Sally} would misuse the data. We assume \emph{Alice} trusts \emph{Sally}  to give her the right answer of \emph{f($x_i$)} though.

To address such concern, one needs cryptographic techniques that allow arbitrary functions to be evaluated on encrypted data. So that \emph{Alice} can send encrypted inputs  \emph{\{$x^1_i$\}} and \emph{Sally} could perform functions on such encrypted data. Fully Homomorphic Encryption (FHE) schemes provide such functionality. They are proven to be practical by Gentry \cite{gentry2009fully}. Current FHE schemes are highly inefficient and still not yet feasible.\cite{vaikuntanathan2011computing}.

Informally, FHE schemes allows one to perform fundamentally \emph{addition,multiplication} over the cipher texts without ever decrypting the inputs. Apart from the \emph{keygen, encrypt, decrypt} algorithms of Public Key Encryption (PKE) systems, FHE schemes consist of an additional algorithm \emph{evaluate}. \emph{Evaluate} method takes input as any arbitrary function \emph{f} in boolean circuit form, encrypted inputs of {f} and executes it over encrypted data using the \emph{Public Key}. The output is still in encrypted form. 

\begin{align} \label{eval}
C_\emph{f}& = \emph{evaluate}_{PK}(\emph{f},C_1,C_2, ...., C_n) 
\end{align}

\subsection{Problem}
Adopting FHE schemes for cloud applications that involves just one user and server is straightforward the data can be encrypted using the \emph{Public key} and outsourced to the server, the server can then perform operations using \emph{Evaluate} method and \emph{Public key} and return the results that are still in encrypted form. The user then decrypts the results using \emph{Private Key}. 

But most of the cloud applications involve multiple users collaboratively utilizing a service.  Encrypting data with their individual \emph{Public keys} would {\bf not} result in encrypted data that can be operated upon by them collectively. 

\subsection{Straw man Solution}
In order to make the encrypted data coherent even after encryptions by multiple users, all the users can have same key pair for encryption and decryption. But this would lose accountability and causes single point of failure. In addition collaboration based on different degrees of trust that needs access control measures cannot be enforced. 

So we precisely need techniques to keep the data coherent even encrypted by multiple users but still each user is accountable for their operations. Additional access control measures also should be enforced easily.

\subsection{Key Contributions} 
\begin{enumerate}
\item We introduce \emph{Complementary Key Pairs} technique that involves two distinct key pairs that complement each other to achieve the desired functionality.
\item  We design protocols for scaling FHE schemes for multi user scenarios and analyze their security using such technique.
\item  We also show how \emph{Server Oblivious - Attribute Based Access Control (SO-ABAC)} can be achieved over encrypted data using such technique.
\end{enumerate}

\section{Prior Art}
 Current popularly published FHE schemes are suited for single user utilizing some cloud service, they are not for multiuser. 

Xiao et all proposed protocols for Multi User systems \cite{xiaoefficient}, that are based on  symmetric Homomorphic Encryption scheme that could evaluate functions only on polynomials. Also their protocols are tightly coupled with their scheme.

Also set of literature exists for achieving multi party computation using threshold homomorphic encryption \cite{cramer2001multiparty},\cite{myers2011threshold} and also multikey homomorphic encryption \cite{lopez2012fly}. These techniques require few of the users to collaborate to decrypt the final result, which is not reasonable to assume especially in the cloud setup.

Non interactive Verifiable Computation \cite{gennaro2010non} is another concept introduced to verify the computations when they are outsourced. Such techniques have been adopted for multiuser scenario too by Choi et all \cite{choimulti}. The multi user protocols introduced here encrypt the same input under multiple identities and stored as a map, this would not scale when the number of users or the input data increases in size.

 Dijk and Juels argue FHE alone cannot solve cloud privacy problems\cite{van2010impossibility}.They prove that, realization of private multi-client scenario is equivalent to impossibility results of program obfuscation. They assume, in multi client scenario the encryption is done with different \emph{Public} keys. They also state that functionality beyond FHE is needed like Access Control over cipher texts and Re-encryption to achieve such private multi-client scenarios. 

  Their impossibility proof makes an assumption that isn�t true in our case. In our protocols, even in multi client scenario the encryption of data is done by single \emph{Public} key and authentication of users is done with different individual key pair. Also we prove (\emph{by construction}) that such additional functionality is purely application of FHE. Our protocols are \emph{counter examples} to their core results.

  Access control over encrypted data is achieved using \emph{Attribute Based Encryption} schemes \cite{goyal2006attribute} \cite{bethencourt2007ciphertext}. Such techniques in combination with \emph{Proxy Reencryption} schemes were adopted to cloud setting \cite{yu2010achieving}. Such adaptations are complex in nature for practical implementation. Also Receiver Oblivious Attribute Based Access Control has been achieved through attribute based encryption \cite{han2012attribute} and this does not fit in cloud setup.
\section{Preliminaries}
  \subsection{Users}
 In all of the below protocols, \emph{Alice, Bob, Carol} are typical Cloud Service Consumers, can be an individual or an enterprise. \emph{Sally} is a typical Cloud Service Provider like Google, Amazon or any other provider. 
 \subsection{Services} The service can be based on any of the \emph{Infrastructure-as-Service (IaaS), Platform-as-Service(PaaS), Software-as-Service(SaaS)} delivery methods \cite {mell2011nist}. 
  \subsection{Adversaries}
 In real world multiple adversaries exist, the cloud platform itself, somebody compromising cloud platform, neighbors sharing the platform can be malicious. Such differentiation is not important though if proper \emph{computational privacy} measures are in place. Abstractly such adversaries can be modeled as  single adversary, \emph{Mallory}. 
 \subsection{Attacks} The goal of \emph{Mallory} is to get unauthorized access to plain text and/or to corrupt the cipher text. \emph{Mallory} can compromise the cloud platform itself or exploit a flaw in the protocol to achieve such goal.
\subsection{Outline}
We start with a trivial protocol for \emph{Alice} to outsource her data \footnote{In communications the abstraction of plain text is message, but in computations data fits well.} to \emph{Sally}  while ensuring \emph{computational privacy}  of the same. We introduce the concept of \emph{Complementary Key Pairs}. We then present protocol for \emph{Alice} to collaboratively work with \emph{Bob} on shared data using such technique. Using this protocol both \emph{Alice,Bob} can execute arbitrary operations on encrypted data. Further we present a protocol that allows \emph{Sally} to enforce oblivious access control over \emph{Alice's} data when multiple users are present. They are analyzed for possible attacks and vulnerabilities.
\subsection{Basic Assumptions}
\begin{itemize}
\item \emph{Alice}, \emph{Bob} and \emph{Sally}  know each other's public keys \footnote{In all the protocols, such keys are given offline or got from trusted server or any other safer means}.
\item \emph{Alice}, \emph{Bob} and \emph{Sally} mutually authenticate each other
\item \emph{Sally} knows legible partners of \emph{Alice} through some initial configuration.
\item All the communications take place on a secure channel
\end{itemize}
 {\bf Note:} Protocols stated here are for illustration purposes, mainly to drive the point. Security is discussed only for the functionality they represent to solve in this context. Conventions followed in protocols are detailed in Table \ref{conventionAll}
\section{Basic Protocol}
In this protocol, we consider the basic scenario where only two parties are involved,  \emph{Alice}  the user and \emph{Sally}  the service provider.
\subsection{Summary}
 \emph{Alice} encrypts the data using her \emph{Public Key} and uploads ( i.e outsources) it to Sally. At any later point of time, \emph{Alice} requests various operations on the encrypted data using API's provided by Sally. \emph{Sally}  performs the operations  on the encrypted data using FHE Scheme's \emph{Evaluate} method and \emph{Public Key} of Alice. \emph{Sally}  returns the results to Alice. \emph{Alice} decrypts the results locally using her \emph{Private Key}.
 
The detail steps of protocol are mentioned in Table \ref{basicProtocol} 
\begin{table}
\caption {Conventions for All Protocols } 
\label{conventionAll}
\renewcommand{\arraystretch}{2.5}
\begin{center}
\begin{tabular} {|c  c|}
\hline
 {A,B,S} & Principal members Alice, Bob, Sally \\
 {$PK_p$,$SK_p$} & is a public,private key pair of principal \emph{p}\\
 {auth-$PK_p$,auth-$SK_p$} & {key pair of any principal \emph{P} for Authentication} \\
 {eval-$PK$,eval-$SK$} & {key pair generated by FHE scheme} \\
 {\emph{func(x)}} & {is an arbitrary function} \\
 {data,x,y} & {users data, input, output respectively to \emph{func}} \\
 {$data^1$,$x^1$, $y^1$} & {are encryptions of data,x,y respectively} \\
 {\emph{encrypt},\emph{decrypt},\emph{evaluate}} & {methods of FHE scheme} \\
 {\emph{sign},\emph{verify}} & {signature and verify operations} \\
 \hline
\end{tabular}
\end{center}
\end{table}
\begin{table}
\caption{Basic Protocol}
\label{basicProtocol}
\renewcommand{\arraystretch}{2.5}
\begin{center}
\begin{tabular} {| c c c c | }
\hline
&\bf{Data Preparation}&&  \\
{1} & {A} & {:=} & {{$data^1$}= $\emph{encrypt}_{PK_a}$(data)} \\
{2} & {A$\quad \rightarrow \quad$S} & {:=} & {$data^1$} \\
\hline
\hline
&\bf{Protocol Steps}&&\\
{1} &{A} & {:=} & {{$x^1$}=$\emph{encrypt}_{PK_a}$(x)} \\
{2} & {A$\quad \rightarrow \quad$S}&{:=} & { (\emph{func},$x^1$)} \\
{3} & {S}&{:=}&{ {$y^1$}= $\emph{evaluate}_{PK_a}$(\emph{func},$x^1$,$data^1$) \eqref{eval}} \\
{4} & {S$\quad \rightarrow \quad$A}& {:=} & {($y^1$)} \\
{5} & {A}& {:=} & {y =  $\emph{decrypt}_{SK_a}$($y^1$)} \\
\hline
\end{tabular} 
\end{center}
\end{table}
 \subsection{Security}
In this simple setup, If cloud platform itself is compromised by \emph{Mallory}, FHE schemes guarantee the \emph{computational privacy} of \emph{Alice's} data. FHE schemes also guarantee that accidental or intentional sharing of encrypted data with other users or service provider's would not leak anything.

 If \emph{Mallory} initiates a request with \emph{Sally}, requesting arbitrary operations on \emph{Alice's} data then the results would still be encrypted. Such requests could be avoided too by incorporating authentication mechanisms using \emph{digital signatures} 
\subsection{Example Use cases}
\begin{itemize}
\item An End user subscribes for SaaS application like Personal Finance Software.
\item An Enterprise outsources their confidential data and applications like financial , human resources records, pay rolls processing to a IaaS provider. 
\end{itemize}
\section{Multiuser Shared Secret Protocol}

In many of use cases, users collaborate with their partners while utilizing a cloud service. Let \emph{Alice, Bob} be two such partners. They both trust each other but don't trust the service provider \emph{Sally}. \emph{Mallory} is an adversary and not a legible partner. The below protocol though is not restricted to two users and scales to any number of users without loss of generality.

Each of these users have their own \emph{Public, Private} key pairs. Encrypting the data with their respective \emph{Public} keys will not result in encrypted data that can be collectively operated upon. At the same time if same key pair is shared among and used by all the users then it would cause single point failure if the keys are compromised. Also there would not be any accountability on who made the change exactly if every one uses same \emph{Public} Keys.
\subsection{Complementary Key Pairs}
 Two distinct types of key pairs are used together through out the protocols. The key pairs \emph{complement} each other to achieve the required functionality.
\begin{enumerate}
\item  Authentication key pair \emph{auth-Private, auth-Public Keys} for Users (\emph{Alice, Bob}) to authenticate
themselves with Server (\emph{Sally}). All the users and server have their own respective \emph{auth pair}.
\item Evaluation key pair \emph{eval-Private, eval-Public keys} for Users to execute arbitrary operations
with Server. All the legible users (partner's of \emph{Alice}) would have one common \emph{Eval} key pair. Server knows \emph{eval-Public} key only.
\end{enumerate}
This is by design principle \emph{Separation of Concerns}. Each key pair is used for separate concern aka functionality.
\subsection{Summary}
Alice, her partners and \emph{Sally}  generate \emph{auth-Private, auth-Public Keys}, this can be either through traditional PKE schemes or FHE. \emph{Alice} also generates \emph{eval-Private, eval-Public keys} using FHE scheme. Alice ensures same \emph{Eval} key pair to be available with all her partners. Such sharing of \emph{Eval} key pair is reasonable and similar to having a pre shared secret in Symmetric key crypto systems.

All the users authenticate themselves using their individual \emph{auth key} pairs while requesting any operations.  The data is encrypted with \emph{eval-Public} key when any of the users \emph{upload/write} it to the cloud and results are decrypted with \emph{eval-Private} key locally at the users. Further the actual execution of the operations are carried out by server using \emph{eval-Public} and \emph{Evaluate} method of FHE scheme.  Since the data is encrypted under same \emph{eval-Public} key by everyone, the encryptions by multiple users are coherent. 

The detail steps of protocol are mentioned in Table \ref{msspProtocol}. \emph{ Bob}'s run of the protocol would be similar, by using his authentication keys, without loss of generality.

\begin{table}
\caption{Multi user Shared Secret Protocol Steps }
\label{msspProtocol}
\renewcommand{\arraystretch}{2.5}
\begin{center}
\begin{tabular} {| c c c c | }
\hline
&\bf{Data Preparation}&&  \\
1 & A &  :=  & $data^1$ =  $\emph{encrypt}_{\emph{eval}\textit{-}PK}(data)$\\
2 & $A \rightarrow S$ & := & $ \emph{sign}_{auth\textit{-}SK_a}(data^1)$\\ 
3 & S &  := & $\emph{verifySig}_{auth-PK_a}(\emph{sign}_{auth\textit{-}SK_a}) $\\
\hline
\hline
&\bf{Protocol Steps}&&\\
{1} & {A} & {:=} & {{$x^1$}=$\emph{encrypt}_{\emph{eval}-PK}$(x)} \\
{2} & {A$\quad \rightarrow \quad$S}&{:=}&{ $\emph{sign}_{\emph{auth}-SK_a}$(\emph{func},$x^1$) } \\
{3}&{S}&{:=} & {$\emph{verifySig}_{\emph{auth}-PK_a}$($\emph{sign}_{auth-SK_a}$)} \\
{4} &{S}&{:=}&{ {$y^1$}= $\emph{evaluate}_{\emph{eval}-PK}$(\emph{func},$x^1$,$data^1$) \eqref{eval}} \\
{5}&{S$\quad \rightarrow \quad$A}&{:=}&{ $\emph{sign}_{\emph{auth}-SK_s}$ ($y^1$)} \\
{6}&{A}&{:=}&{ $\emph{verifySig}_{\emph{auth}-PK_s}$($\emph{sign}_{\emph{auth}-SK_s}$) } \\
{7}&{A}&{:=}&{y =  $\emph{decrypt}_{\emph{eval}-SK}$($y^1$)} \\
\hline
\end{tabular} 
\end{center}
\end{table}
\subsection{Security}

All the users use their own \emph{auth-keys} used for authentication purposes. The actual encryption of the data is done using \emph{eval-Public} key. Since the encryption is done under single key by all the users, its easy for the server to run \emph{Evaluate} method of FHE scheme. Similarly all the users (except the server) have \emph{eval-Private} key for decrypting the results locally.

 \emph{Mallory} cannot request for arbitrary operations over data shared by \emph{Alice, Bob}, such requests would be dropped while verifying the \emph{signature} since \emph{Sally} knows the legible users. If \emph{Mallory} some how compromises any user's \emph{auth-Private} key, she may successfully impersonate such user and request arbitrary operations but she may not be able to \emph{decrypt} the results as that would need \emph{eval-Private} key. Accidental compromise of  \emph{eval-Private} key alone will not allow \emph{Mallory} to perform operations on the data since authentication is done by \emph{auth} key pair.

The security of the system depends though on adequate measures for safer key distribution and authentication mechanisms.
 
\subsection{Example Use cases}

In the below cases \emph{Eval  key pair} can be shared among \emph{Alice} and his partners.
\begin{itemize}
\item An Enterprise has its branch offices located at various different locations. All the branches collectively need to work over data and applications with remote cloud Server.
\item An End user wants to store his personal data with cloud Server. It is very likely to share confidential data with few other users (like family members).
\end{itemize}

\section{Multi-user Constrained Secret Protocol}

A much richer context is where \emph{Alice} authorizes her partner users with different access control privileges. In many of real world scenarios, a user would authorize partners to obtain results of only certain operations. Also \emph{Alice} would like to have control on \emph{addition, deletion, revocation etc } on the partner users apart from authorizing granular access control.

Informally, Access control is \emph{authorizing}  subjects (users, applications or processes) with certain levels of access to various resources (files, devices etc ) and enforcing such \emph{authorization} when the resources are accessed. A generalized model of all the traditional ones is Attribute Based Access Control (ABAC)\cite{jin2012unified}. ABAC is an excellent choice for web services like cloud computing due its feature richness\cite{yuan2005abac}.

We give a quick introduction to ABAC model. We continue to use the \emph{complementary key pairs} technique with additional restrictions for access control. Although all the legible users have the \emph{eval-Private} key to decrypt the results they are further \emph{constrained} to perform only certain operations, thus the name \emph{constrained secret protocol}. We first show how \emph{access control} can be achieved on encrypted data using FHE schemes and then use those techniques in the protocol.
\subsection{Attribute Based Access Control (ABAC)}

In ABAC model, permissions are defined based any characteristics known as \emph{attributes}. The attributes for \emph{subjects} and \emph{resources} can be identifiers, names, title etc. ABAC also has special kind of \emph{environmental} attributes like current date, security levels etc.  A \emph{policy rule} defined as a boolean function of attributes to decide if a subject \emph{s} can access a resource \emph{r} in a environment {e} as below. \\
\begin{align} \label{canaccess}
 \emph{canAccess}& = \emph{f}(ATTR(s),ATTR(r),ATTR(e)) 
\end{align}

An \emph{Administrator} defines such attributes and also decides such policy rules \footnote{In reality attribute administrator and policy administrator can be different too}. They are stored in a \emph{Policy Rule Base(PRB)}. A \emph{Policy Enforcement Point (PEP)} is responsible for receiving requests for authorization decisions and enforcing them. PEP in turn takes help of \emph{Policy Decision Point (PDP)} to evaluate the validity of authorization request by processing Policy Rules.

In our current cloud setup, The subjects are \emph{Alice,Bob}. Encrypted data fine grained as files or database records are resources. \emph{Alice} as the data owner takes the role of an administrator, defines on the attributes for partner users and resources. \emph{Alice} also defines the policy rules she wants to grant as privileges for her partners. \emph{Alice} outsources such PRB to \emph{Sally} the server. \emph{Sally} acts as both Policy Enforcement and Decision Point. Also \emph{add,delete,revoke} of partner users can be achieved using the PRB. For example, absence of privileges for a particular user in PRB can be considered as the user is deleted or revoked.

In Cloud setting, Policy Rule Base(PRB) in plain form, at the server, can \emph{leak} information on the underlying data. For example, if a policy rule as "\emph{Carol} can \emph{write} to a file \emph{xyz.jpg}", would leak information. For this reason, the PRB also need to be encrypted so that server is \emph{oblivious} of such details.

\emph{Server Oblivious - Attribute Based Access Control(SO-ABAC)} can be achieved using FHE schemes together with \emph{Complementary Key Pairs} technique introduced in previous protocol. Attributes can be defined for users identified by their respective \emph{auth-Public Keys}. The resource identifiers can be encrypted using \emph{eval-Public Key} of FHE scheme. Enforcement of such access privilege can be done, obliviously a.k.a blind-foldedly, using \emph{Evaluate} method for any user requesting access for a resource.
\subsection{Summary}
The current protocol is basically an extension of previous protocol. Additional conventions followed are captured in Table \ref{conventionAddl}
\begin{table}
\caption {Additional Conventions} 
\label{conventionAddl}
\renewcommand{\arraystretch}{2.5}
\begin{center}
\begin{tabular} {|c  c|}
\hline
{PRB} & {Policy Rule Base} \\
{$PRB^1$} & {an encrypted PRB} \\
{s,r,e} & {subject, resource, environment} \\
{ATTR(a)} & {attribute for s ,r or e} \\
{$ATTR^1$(a)} & {attribute encrypted using eval-PK} \\
{\emph{verifyAccess}}&{verify access control method used by Sally} \\
 \hline
\end{tabular}
\end{center}
\end{table}

In addition to previous protocol's assumptions, \emph{Sally} knows that \emph{Alice} is also the administrator and has the permissions to  manage the PRB. Initial data preparation would be same as previous protocol. In addition, \emph{Alice} defines PRB. \emph{Sally} in turn grants access to users as per encrypted PRB.

 Alice as Administrator configures PRB and subsequently \emph{Bob} as partner user requests operations. 
  
  When an user requests certain operations, he/she needs to send certain attributes. The attributes of a user \emph{Bob} used for identification is {\emph{auth}-$PK_b$}, there can be other attributes too. The attributes for resource and environment depends on the operation being requested lets assume they are \emph{r,e}. All the attributes and PRB are encrypted using \emph{eval-Public} key.
\begin{subequations}\label{eattr}
\begin{align}
&ATTR^1(a_1 ... a_n) = \emph{encrypt}_{\emph{eval}\textit{-}PK}(a_1 .... a_n) \quad a_i \in \emph{s, r, e}\label{at3}\\
&ATTR(s) = \emph{auth}\textit{-}PK_b \label{at1} \\ 
&ATTR^1(s) = \emph{encrypt}_{\emph{eval}\textit{-}PK}(\emph{auth}\textit{-}PK_b)\label{at2}\\
&PRB^1 = \emph{encrypt}_{\emph{eval}\textit{-}PK}(PRB)\label{pbr}
\end{align}
\end{subequations}

 Policy enforcement decision can be done at the server using the \emph{Evaluate} method of FHE scheme as shown below, this is later used in the protocol. Assume \emph{canAccess} \eqref{canaccess} is a boolean circuit of ABAC model.
\begin{multline} \label{access}
\emph{verifyAccess}_{\emph{eval}\textit{-}PK}(\emph{func}, ATTR^1(a_1 .. a_n)) \\ = \emph{evaluate}_{\emph{eval}\textit{-}PK}(\emph{canAccess},\emph{func},ATTR^1(a_1 .. a_n),PRB^1)
\end{multline}

The detail steps of protocol are mentioned in Table  \ref{mcspProtocol}.  \emph{Alice}'s run of the protocol, as user, would be similar, by using his authentication keys ,  without loss of generality.

\begin{table}
\caption{Multi user Constrained Secret Protocol }
\label{mcspProtocol}
\renewcommand{\arraystretch}{2.5}
\begin{center}
\begin{tabular} {| c c c c | }
\hline
&\bf{PRB Preparation}&&  \\
{1}&{A}&{:=}&{{$PRB^1$} \eqref{pbr}} \\
{2}&{A$\quad \rightarrow \quad$S}&{:=}&{$\emph{sign}_{auth-SK_a}$($PRB^1$)} \\
{3}&{S}&{:=}&{$\emph{verifySig}_{auth-PK_a}$($\emph{sign}_{auth-SK_a}$)} \\
\hline
\hline
&\bf{Protocol Steps}&&  \\
{1}&{A}&{:=}&{{$x^1$}=$\emph{encrypt}_{\emph{eval}-PK}$(x)} \\
{2}&{A$\quad \rightarrow \quad$S}&{:=}&{ $\emph{sign}_{\emph{auth}-SK_b}$(\emph{func},$x^1$,$ATTR^1$({$a_1$...$a_n$)}) \eqref{at3} } \\
{3}&{S}&{:=}&{$\emph{verifySig}_{\emph{auth}-PK_b}$($\emph{sign}_{auth-SK_b}$)} \\
{4}&{S}&{:=}&{$\emph{verifyAccess}_{\emph{eval}-PK}$(\emph{func},$ATTR^1$({$a_1$...$a_n$)}) \eqref{access}} \\
{5}&{S}&{:=}&{ {$y^1$}= $\emph{evaluate}_{\emph{eval}-PK}$(\emph{func},$x^1$,$data^1$) \eqref{eval}} \\
{6}&{S$\quad \rightarrow \quad$A}&{:=}&{ $\emph{sign}_{\emph{auth}-SK_s}$ ($y^1$)} \\
{7}&{A}&{:=}&{ $\emph{verify}_{\emph{auth}-PK_s}$($\emph{sign}_{\emph{auth}-SK_s}$) } \\
{8}&{A}&{:=}&{y =  $\emph{decrypt}_{\emph{eval}-SK}$($y^1$)} \\
\hline
\end{tabular} 
\end{center}
\end{table}
\subsection{Rekeying, Revocation and ReEncryption}

If \emph{Bob}'s \emph{auth key} pair is compromised. He generates a new key pair and sends the \emph{auth-Public} key to \emph{Bob}. \emph{Alice} needs to update the PRB with identification attribute of \emph{Bob} containing the new public key.
If \emph{Alice} needs to revoke a particular user, she creates and encrypts such a policy rule and sends a update PRB request to \emph{Sally}.

If \emph{eval-Private} key is compromised, \emph{Alice} generates a new \emph{eval`} key  pair and distributes it to all the partner users. She lets \emph{Sally} know the new public key \emph{eval-PK`} key and requests a re-encryption of data. \emph{Sally} executes the \emph{evaluate} method with the encryption circuit and new public key as below 
\begin{equation*}
reEncryption :  \emph{evaluate}_{\emph{eval}-PK}(\emph{encrypt},\emph{eval-PK`},data^1)
\end{equation*}
\subsection{Security}

If \emph{Mallory} is not a legible partner but tries to access resources that are forbidden, such requests are first thwarted while verifying signature itself, since its assumed \emph{Sally} knows all legible users through some initial configuration.
If \emph{Mallory} is legible partner but tries to access resources that are forbidden, such requests are thwarted while verifying the access control privileges. If \emph{Mallory} tries to overwrite the PRB itself, such requests are thwarted too since such administrative operations are allowed by \emph{Sally} only from \emph{Alice}. 

A risk exists though, If \emph{Mallory} compromises the cloud platform itself and corrupts the PRB or bypasses the access check, this could result in a complete denial of service. But such \emph{Single Point of Failure} is inherent to cloud computing model itself, there is a total denial of service to user due to poor internet connectivity or if cloud service is down (either due to technical problems or attacks).
\subsection{Example Use cases}
   \begin{itemize}
\item \emph{Privacy of Health Records} Hospitals can decide access control on  their data based on who their partners are for example insurance, doctors, researchers etc. Server can be completely blind about the data being processed and also the access privileges for resources.
\item \emph{Private Social Networking} Each user can authorize, who (friends, family, everyone etc) can access their data (like photos, blogs etc) and this can be done without the server ever knowing it.
\end{itemize}
\section{Future Work}
 We have assumed, in our protocols, the cloud server would be honest in performing the computations correctly, in real world such assumptions cannot be made. Accidentally or malicious execution of arbitrary computations over encrypted data might effect the results and/or render data useless too. For this reason, \emph{verifiability of the computations} is very important. Current verifiability techniques are nascent \cite{gennaro2010non}\cite{benabbas2011verifiable}\cite{chung2010improved}. Enhancing such techniques and tailoring them into the above protocols would be for further work. Rigorous proofs of security protocols for computational privacy can be achieved when \emph{verifiability} of computations is integrated with them.
 
 Current FHE schemes are highly inefficient and theoretical to conduct experiments and provide results. Also Cloud Computational Privacy protocols being nascent, there aren't any theoretical frameworks to best capture their security requirements and conduct rigorous analysis. The current popularly used Universal Composability framework has its own limitations \cite{canetti2003limitations} to be applied directly to cloud protocols.
 \section{Conclusions}
   We introduced \emph{complementary key pairs} technique for multi user scenarios.Three protocols which cater to gamut of use cases of cloud computing identified by the community \cite {computing2010cloud} are presented in our work. Also we showed that FHE schemes for single user can be scaled to multi user scenario too and no special variants of FHE schemes are required for the same. These protocols make no assumptions on underlying schemes, this gives them flexibility of adapting them to any FHE schemes. We also showed how oblivious (blind-folded) \emph{access control} over encrypted data can be achieved using FHE schemes. 
\section{Acknowledgments}
I would like to thank Cisco Systems for  supporting this work. I would like to thank my research advisor Dr V.N. Muralidhara for all the insightful discussions and brainstorming. I would like to thank my colleague and Technical leader Scott Fluhrer, a brilliant cryptographer in Cisco for providing valuable suggestions for this paper. I would like to thank Cisco Fellow Dr David McGrew for all the opportunities he created for me to learn cryptography and cloud security in detail.  I would also like to thank \emph{crypto.stackexchange} community for having many insightful discussions around this topic.

\bibliographystyle{IEEEtran}	
\bibliography{IEEEabrv,conf}
\end{document}